\documentclass[aps,prl,preprint,groupedaddress]{revtex4-2}

\usepackage[pdftex]{graphicx}

\usepackage{amsmath}
\usepackage{amssymb}

\usepackage{empheq}

\usepackage{tabularx}
\usepackage{footmisc}

\begin{document}


\title{In situ characterization of laser-induced strong field ionization phenomena}


\author{Noam Shlomo and Eugene Frumker}
\email[]{efrumker@bgu.ac.il}
\affiliation{Department of Physics, Ben-Gurion University of the Negev, Beer-Sheva 84105, Israel}

\date{\today}

\begin{abstract}
Accurately characterizing the intensity and duration of strong-field femtosecond pulses within the interaction volume is crucial for attosecond science. However,  this remains a major bottleneck, limiting accuracy of the strong-field, and in particular, high harmonic generation experiments.
We present a novel scheme for the in situ  measurement and control of spatially resolved strong-field femtosecond pulse intensity and duration within the interaction focal region. Our approach combines conjugate focal imaging with in situ ion measurements using gas densities pertinent to attosecond science experiments. Independent measurements in helium and argon, accompanied by a fitting to a strong field ionization dynamic model, yield accurate and consistent results across a wide range of gas densities and underscores the significance of double ionization, as well as barrier suppression ionization. 
Direct spatially resolved characterization of the driving laser is a critical step towards resolving the averaging problem in the interaction volume, paving the way for more accurate and reliable attosecond experiments.
\end{abstract}


\maketitle

\section{Introduction}

Ultrafast laser-induced strong field phenomena play a pivotal role across a broad spectrum of fundamental areas within modern physics. 
For instance, the first step in the three-step model of high harmonic generation (HHG) \cite{Corkum_3step_PRL1993, Frumker_orientation_HHG_PRL2012}, a cornerstone of attosecond science, relies on strong field-driven tunnel ionization. 
Similarly, powerful techniques like 'attoclock' employ this phenomenon to probe intricate details of quantum tunnelling \cite{Eckle_tunneling_time_delay_Science_2008, Sainadh_Litvinyuk_tunnelling_time_in_H_Nature_2019}. Strong field induced Coulomb explosion, fragmentation, tunneling and multiphoton ionization in the velocity map imaging (VMI) \cite{Eppink_VMI_1997} and cold target recoil ion momentum spectroscopy (COLTRIMS) \cite{Ullrich_Recoil_RepProgPhys_2003} experiments are used to study the variety of most fundamental physical and chemical processes in atoms and molecules \cite{Akagi_tunnel_ion_Science2009, Weber_correlated_nature_2000}. 
In the wake-field acceleration  experiments \cite{Tajima_Laser_accelerator_PRL_1979, Malka_wake_field_Science_2002}, the strong laser field allows to accelerate particles with a record gradients, enabling the construction of high performance accelerators of much smaller size than conventional ones or even at the photonic chip scale \cite{Peralta_acceleration_Nature2013, Breuer_acceleration_PRL_2013, Shiloh_Electron_Nature_2021}.

Accurate measurement, optimization and control of laser intensity distribution and pulse duration in the interaction region are critically important for most of the strong field-driven experiments.


Upon initial examination, the problem may appear straightforward: to directly  measure or estimate fundamental laser pulse characteristics, such as energy, width, and duration, and subsequently deduce the peak intensity. However, it has long been recognized and experimentally validated  \cite{Huillier_Multiply_charged_ions_1dot06_JPhysB_1983, Alnaser_Laser_peak_intensity_calibration_COLTRIMS_PRA_2004, Frumker_orientation_HHG_PRL2012} that such a seemingly simple approach usually results in unacceptably large errors, frequently exceeding 50\%, which can significantly impact the interpretation of experimental results. Several factors can contribute to such errors, including imperfect laser beam quality, aberrations in the focusing optics, distortion of ultrafast pulses due to spatial chirp (e.g., caused by misalignment of the compressor grating), pulse dispersion in the generation medium, and more.  

Furthermore, high-power laser systems often are prone to temporal fluctuations in pulse intensity and variations in their spatial and temporal profiles, compounding the issue due to the  highly nonlinear nature of strong-field phenomena. Therefore, a reliable in situ approach to measure and control these parameters in the focal volume becomes critically important for all areas involving strong-field laser physics.

To address this challenge, various methodologies have been reported. Typically, intricate, specialized, and costly techniques have been employed to achieve this objective, including mass spectrometry \cite{Lai_SFI_experimentalPRA_2017}, velocity map imaging (VMI) \cite{Smeenk_Precise_in_situ_measurement_of_laser_pulse_intensity_VMI_OE_2011}, and cold target recoil-ion momentum spectroscopy (COLTRIMS) \cite{Alnaser_Laser_peak_intensity_calibration_COLTRIMS_PRA_2004}. Furthermore, the utilization of these approaches necessitates very low gas densities, generally incompatible with HHG experiments, and often extremely low transverse temperatures, which further complicate the experimental setup, requiring larger system dimensions. 

For high gas density measurements, particularly critical in strong-field physics, such as attosecond science, the estimation of laser peak intensity has often relied on the high harmonic generation (HHG) cutoff \cite{Itatani_tomog_Nature2004, frumker_wavepacket_PRL2012}. However, in practice, determining the cutoff is nontrivial, as contrary to an idealized and clearly defined cutoff, a series of low-intensity harmonics frequently appear in the experimentally measured HHG spectrum (\cite{Frumker_sword_OL2009, Frumker_orientation_HHG_PRL2012}). As a result, determining the cutoff harmonic order is somewhat subjective. Furthermore, it has been demonstrated that macroscopic effects, arising from averaging and phase matching, can distort the single-atom HHG cutoff response \cite{Huillie_HHG_cut_off_PRA_1993}, leading to inaccurate intensity estimates and a lack of access to pulse duration estimation within the interaction region.

In this study, we introduce a novel in situ measurement and control approach that solves this long-standing critical problem, opening the door to unprecedented accuracy in strong-field, in particular, attosecond science experiments.
By conducting independent measurements in helium (He), the gas with a highest known ionization potential, $I_p = 24.6$ eV, and argon (Ar), $I_p = 15.76$ eV, \cite{NIST_ASD_database_2017} with exact the same beam parameters in the focal volume, we have identified indelible and meaningful signatures of double ionization, and barrier-suppression ionization (BSI).


Furthermore, we realized a dynamic theoretical model that incorporates single ionization, sequential double ionization (SDI), non-sequential double ionization (NSDI), non-sequential excitation (NSE), recollision-induced excitation with subsequent ionization (RESI), and BSI phenomena for strong field ionization. This model demonstrates excellent agreement and consistency with our experimental results. Importantly, while demonstrated here using argon as a test case, this approach seamlessly extends to the experiments with other atomic gases, including neon (Ne), krypton (Kr), and xenon (Xe).



\begin{figure}
    \begin{center}
	\includegraphics[width=0.8\linewidth]{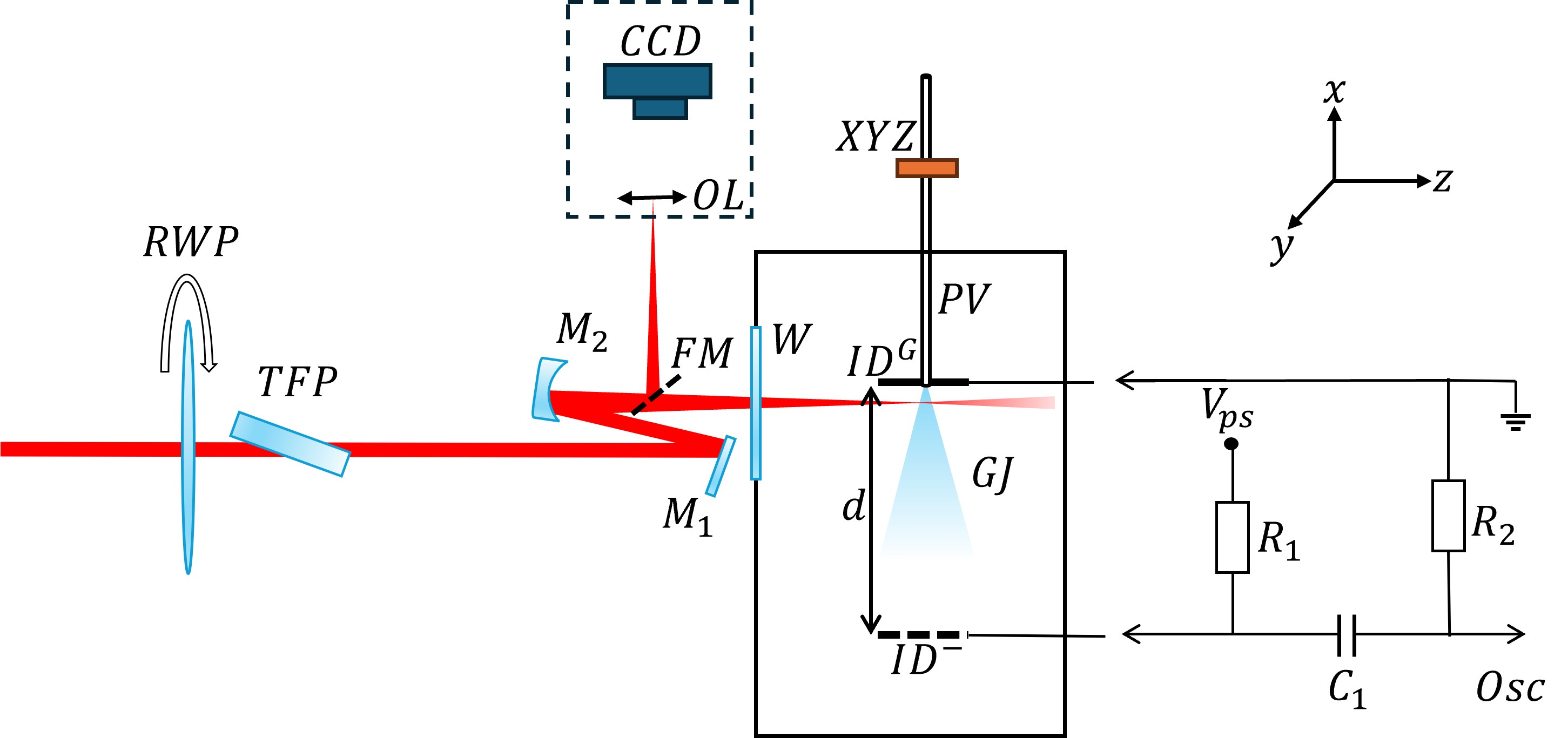}
    \caption{\label{Fig_SetUp}  Schematic of the experimental setup. The ion detector mesh $\textrm{ID}^\textrm{-}$ is connected to the negative voltage of the high voltage power supply ($V_{PS}$)  through the shunt resistor ($R_1= 1M\Omega$). The output is monitored by an oscilloscope (OSC) connected to the circuit via a DC decoupling capacitor ($C_1 = 100 pF$) and a  resistor ($R_2=10 M\Omega$). The input capacitance of the scope (15 pf) and input impedance ($1M\Omega$) are not shown in the figure but are accounted for in the calculations.} 
    \end{center}
\end{figure}	

\section{Results}
\subsection{Ion detection scheme}

The schematic of our experimental setup is shown in Fig. \ref{Fig_SetUp}. A linearly polarized laser pulse from a Ti:sapphire amplifier, centered around $\lambda = 800$ nm with a pulse duration of 30 fs (measured at the laser output), was focused by a spherical f=500 mm focusing mirror M2 through the vacuum chamber window (W) onto a gas jet (GJ).  To maintain precise control over the pulse energy, a motorized $\lambda/2$ waveplate (RWP) and a thin-film polarizer (TFP) were employed. The focal plane on the gas jet was monitored via an image-conjugated microscope, which comprised a beam-picking mirror (FM), an objective lens (OL), and a CCD imager.

  The ion detector  \cite{shiner_wavelength_scaling_2009, frumker_wavepacket_PRL2012, Tchulov_LaserStrongFieldTom_ScRep_2017} is composed of a stainless steel grounded plate ($\textrm{ID}^\textrm{G}$) connected mechanically and galvanically to the body of the pulsed valve ,  and the metal mesh ($\textrm{ID}^\textrm{-}$) attached to the ($\textrm{ID}^\textrm{G}$) with dielectric rods and connected to the negative  high-voltage, $V_{PS}$ via a simple passive electronic circuit shown in Fig. \ref{Fig_SetUp}. In this experiment, we employed a pulsed valve (Parker series 9 solenoid valve) that is widely recognized for its versatility and widespread use in numerous strong-field experiments. It is important to note that our approach is not limited to this specific valve and can seamlessly adapt to other gas jet sources as well. We applied a voltage of $V_{PS}=-500$ V and set the distance between the grounded plate and the mesh to d=80 mm. Both the grounded plate and the mesh are circular, each with an 80 mm diameter.

  The inclusion of the mesh, instead of solid component, is instrumental in minimizing the gas flow resistance, thereby enabling efficient evacuation of the residual gas by a vacuum turbomolecular pump located beneath the vacuum chamber.
  The pulsed gas valve was installed on a micron-precision (XYZ) vacuum positioner, ensuring that the laser beam intersected the jet at the center, approximately $100 \mu \textrm{m}$ below the injection flange. This intentional positioning was chosen for optimal conditions for attosecond science  experiments (HHG), facilitating in situ measurements of HHG \cite{frumker_wavepacket_PRL2012, Frumker_orientation_HHG_PRL2012}.
  This straightforward  construction not only enhances the versatility of the ion detector but also simplifies its deployment across a wide range of experimental configurations. 
  Consequently, our approach can be easily applied in a diverse array of strong-field experiments.
    
    \begin{figure}
    \begin{center}
	\includegraphics[width=0.7\linewidth]{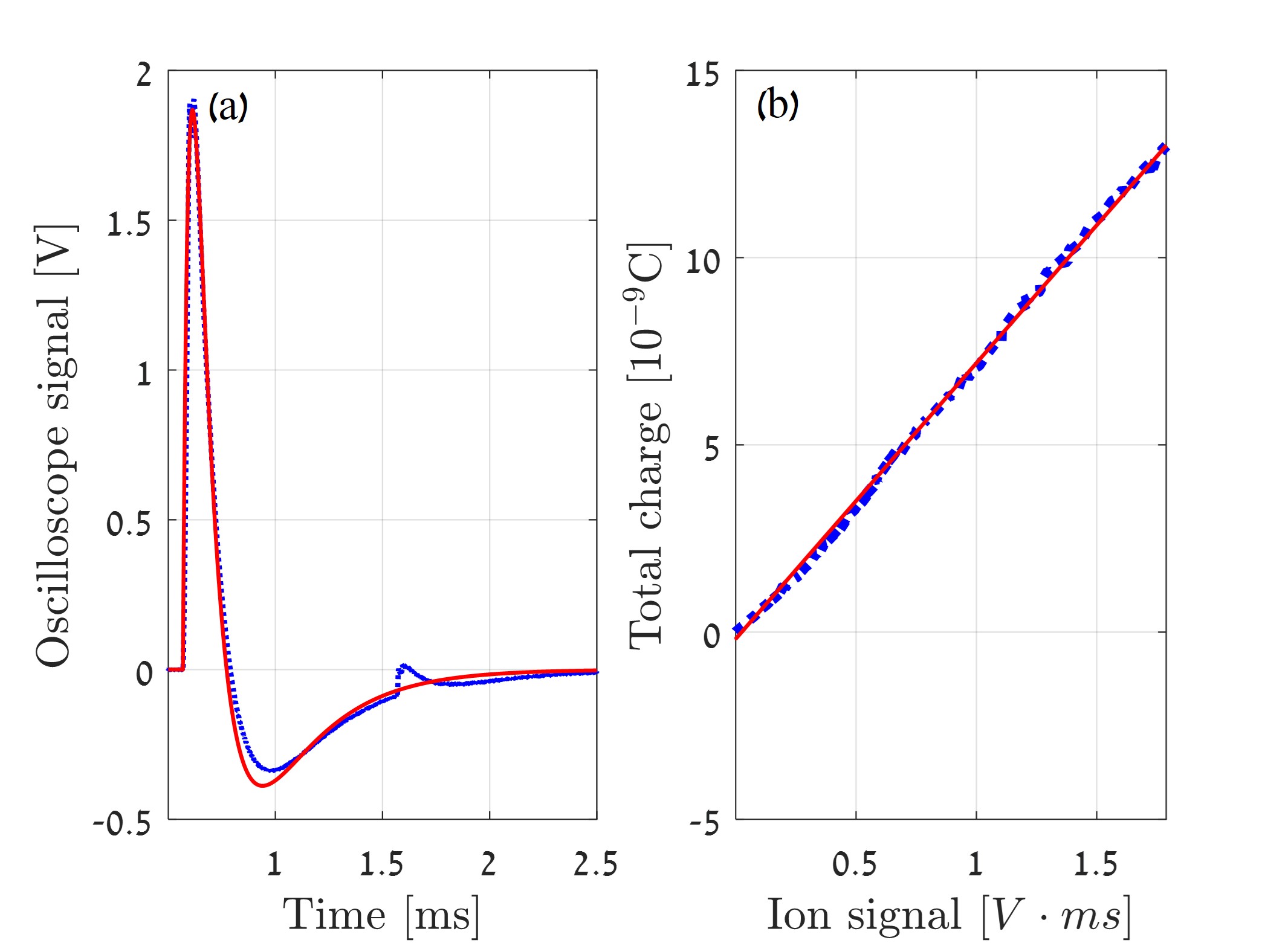}
    \caption{\label{Fig_ion_signal}  (a) Example of measured (dotted blue curve) vs. calculated (solid red curve) ion signal temporal evolution for argon jet. The laser pulse energy was 0.6mJ, and the gas backpressure was 7.5 Bar. The best fitted parameters were found $R_{ID}(t_0)=13.27 M \Omega $  and $\tau_{ID}=30.74 \mu s$  with goodness of fit $R^2=0.98$. (b) Total charge (integrated current, modeled in LTspice and shown by the dotted blue curve) vs. measured ion signal. The goodness of linear fit (solid red line) $R^2=0.999$. 
  }
    \end{center}
\end{figure}

     When the laser pulse interacts with the gas jet, the gas molecules and atoms undergo ionization within the pulse's relatively short duration, typically on the order of tens of femtoseconds, through the strong-field ionization mechanisms. As a result, the gas medium's resistance drops, enabling current to flow between the ground plate and the mesh, recorded as the voltage evolution on an oscilloscope as illustrated in Fig. \ref{Fig_SetUp}. Utilizing a professional analog electronic circuit simulator LTspice, we performed the calculations 
      to determine the transient response of the circuit, and fit it to the experimentally measured voltage trace, with a typical result shown in Fig. \ref{Fig_ion_signal} (a). 
      
      The laser ionized gas was modeled as a time-dependent resistor $R_{ID}(t)=R_{ID}(t_0)\exp[(t-t_0)/\tau_{ID}]$ in parallel with a capacitor, representing the  ion detector geometry. Transmission lines with defined impedances and lengths were included to account for cable effects. The initial resistance $R_{ID}(t_0)$, ionizing pulse arrival time $t_0$, and the decay constant $\tau_{ID}$ were fit to experimentally measured  data.

     The minor "bump" observed 1 millisecond after the primary pulse results from the ionization of residual gas induced by a successive laser pulse from our 1 kHz amplified laser system. 
      We systematically varied the laser intensity by adjusting the Rotating Waveplate (RWP) in Fig. \ref{Fig_SetUp}, and recorded the voltage trace for each intensity setting. 
      For each measured voltage trace, we used the LTspice to compute the total charge, which corresponds to the integral of the current between the ground plate and the mesh, and confirm a linear dependence between the ion signal and the total charge as shown in Fig. \ref{Fig_ion_signal} (b). The term "ion signal" can refer to either the maximum voltage measured on the oscilloscope trace or the integral of the voltage trace (as used in our analysis), both result in the linear dependence between the ion signal and the total charge.

    \begin{figure}
    \begin{center}
	\includegraphics[width=0.7\linewidth]{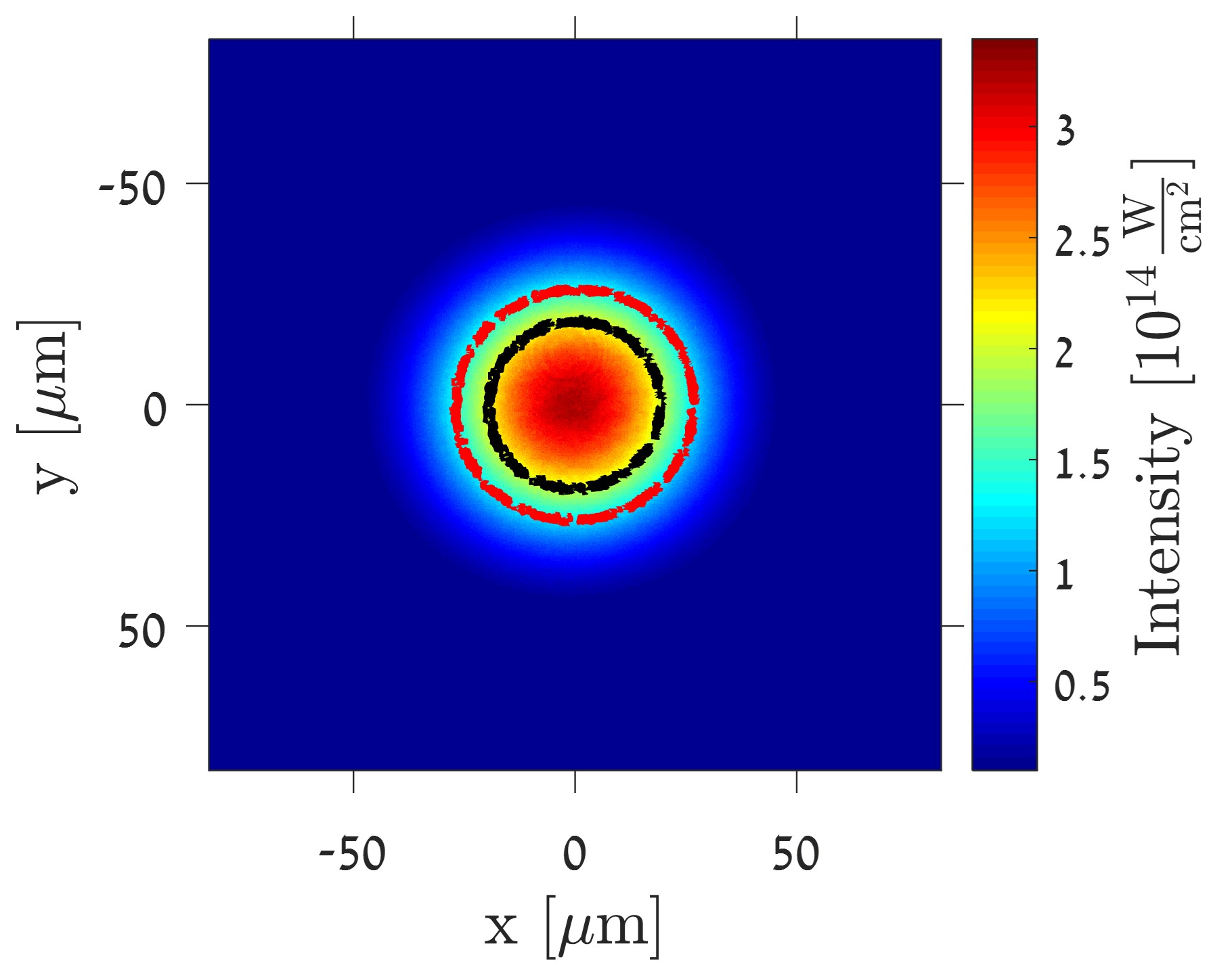}
    \caption{\label{Fig_BeamFocusProfile}   Measured intensity distribution of the beam at the focal plane. The inner black  and the outer red lines represent the $\gamma=1$ contours for helium  and argon gases respectively, emblematically illustrating the regions of tunneling (inner part) and multiphoton (outer part) regimes.
The intensity was calibrated using energy per pulse of 0.4mJ and $\tau_{FWHM} = 43$ fs. The spatial intensity distribution in the focal plane was imaged using a microscope as illustrated in Fig. \ref{Fig_SetUp}.  
  }
    \end{center}
\end{figure}

The measured spatial intensity distribution of the beam  in the focus is shown in Fig \ref{Fig_BeamFocusProfile}. The input beam has been apertured down to 9.25 mm  before the focusing mirror (M2) with an iris to improve the beam quality and to extend effective Raleigh range in the focus. By scanning the image-conjugated microscope (depicted as a dashed box in Fig. \ref{Fig_SetUp} ), we validated that the beam width and shape do not change within the measured 0.4 mm extension (FWHM) of the gas jet.

\subsection{Strong Field Ionization Dynamics}


Before delving into intricacies of strong-field ionization processes, it is essential to emphasize the clear qualitative distinctions between the perturbative and strong-field regimes of nonlinear responses of matter to a laser field.  

 In the perturbative regime, at relatively low intensities (typically $\leq10^{13}$ W/cm$^2$), the interaction between the laser's electric field and the electrons within an atom or molecule is significantly weaker than the Coulomb interaction between the electrons and the nucleus. As a result, the material's polarizability can be expanded as a converging power series in the laser electric field strength \cite{Franken_SHG_PRL_1961, Armstrong_Interactions_Nonlinear_PRL_1962}. Many hallmark nonlinear optical phenomena arise in this regime \cite{Boyd_NLO}, such as second- and third-harmonic generation \cite{Franken_SHG_PRL_1961, Terhune_THG_PRL_1962}, parametric down-conversion \cite{Harris_PDC_PRL_1967}, self-phase modulation \cite{Shimizu_SPM_Liquids_PRL_1967, Alfano_SPM_solids_PRL_1970}, and supercontinuum generation \cite{Alfano_supercontinuum_generation_PRL_1970}, among others. 

In contrast, in the strong-field regime, where the optical field strength becomes comparable to the Coulomb interaction between the electron and nucleus (typically $\geq10^{14}$ W/cm$^2$), the polarizability power series expansion no longer converges. This regime is dominated by ionization processes, which play a pivotal role in our measurements.
 Single-electron ionization of neutral atoms or ions represents the most fundamental aspect of strong-field ionization and has been extensively studied for several decades \cite{Keldysh_ionization_JETP1965, Perelomov_PPT_JETP1966, Ammosov_ADK_JETP_1986, Eckle_tunneling_time_delay_Science_2008, Sainadh_Litvinyuk_tunnelling_time_in_H_Nature_2019}.  Keldysh's seminal contribution \cite{Keldysh_ionization_JETP1965} laid the foundation for our understanding of strong-field ionization in the so-called 'transparency region', where the ionization potential, $I_p$, exceeds the energy of the incident laser field, $\hbar \omega_L$. 

The nature of the ionization process is characterized by the adiabaticity or Keldysh parameter, denoted as $\gamma$. This parameter is defined as $\gamma =  2 \pi \frac{T_{\textrm{tunnel}}}{T_L}$, where $T_L = \frac{2\pi}{\omega_L}$ represents the laser field's period, and $T_{\textrm{tunnel}}$ is referred to as the "tunneling time."
The tunneling time \cite{Keldysh_ionization_JETP1965} is defined as a ratio of twice the length of the potential barrier, $\textrm{L}_\textrm{barrier}$ divided by a classical velocity, $\textrm{v}_{\textrm{e}}$, of an electron moving under the barrier. The electron's velocity is given by the virial theorem: $\textrm{v}_{\textrm{e}} = \sqrt{2I_p / m_e}$, where $m_e$ is the electron's mass. Assuming a point-like potential for electron-nuclear interaction and approximating the total potential with a triangular barrier solely created by the laser field, one gets $\textrm{L}_\textrm{barrier}= I_p /{e E_L}$. Consequently, for the tunneling time, we get $T_{\textrm{tunnel}}=\frac{2\textrm{L}_\textrm{barrier}}{{\textrm{v}}_{\textrm{e}}}=\frac{\sqrt{2I_p m_e}}{ e E_L}$, and for the Keldysh parameter, we obtain:
\begin{equation}
\label{Keldysh_parameter}
\gamma  = 2 \pi \frac{\sqrt{2I_p m_e}}{ e E_L T_L}.
\end{equation}

For a given laser frequency,  a weaker laser field, $E_L$, results in a longer potential barrier length, $\textrm{L}_\textrm{barrier}$, leading to an increase in the Keldysh parameter, $\gamma$.  In the limit of $\gamma \gg 1$ or $E_L \ll \sqrt{2I_p m_e} \omega_L/e$, the laser field undergoes multiple oscillations during the tunneling time, defining the multiphoton regime. In this regime, the ionization rate, denoted as $W_i$, exhibits a power-law dependence on the electric field, a well-known feature in perturbative non-linear optics \cite{Boyd_NLO}: $W_i\propto E_L^{2K}$, with $K=\lfloor I_p/\omega_L+1 \rfloor$ representing the minimum number of absorbed photons required to conserve energy, known as multiquantumness. Conversely, in the limit of  $\gamma \ll 1$ or $E_L \gg \sqrt{2I_p m_e} \omega_L /e$, the driving laser field experiences minimal change during the tunneling time, 
 defining the well-known  tunnelling regime with ionization rate for hydrogen atom given in Gaussian units \cite{Landau_QM_V3} as $W_i = \frac{4 m_e ^3 e^9}{E_L \hbar ^ 7} \exp[-2E_a/(3E_L)]$, where $E_a={m_e}^2 e^5/\hbar^4$ is the electrical field experienced by an electron situated at the Bohr radius $a_0=\frac{\hbar^2}{m_e e^2}$ in the hydrogen atom.

It is essential to emphasize that in our experiment, as illustrated by the measured profile in Fig. \ref{Fig_BeamFocusProfile}, the intensity varies across the beam so that at the central region of the beam, ionization may predominantly occurs in the tunneling regime, while as we move away from the center, the field strength gradually decreases, transitioning towards the multiphoton ionization regime. This intermediate regime is particularly rich and interesting as it contains signatures of both tunnelling and multiphoton ionization mechanisms. Helium's higher ionization potential restricts the dominant tunneling ionization to a smaller area compared to argon.


Shortly after seminal Keldysh work, Perelomov, Popov and Terentev in their work \cite{Perelomov_PPT_JETP1966, Perelomov_PPT2_JETP1967, Perelomov_PPT3_JETP1967} addressed the effect of Coulomb interaction on ionization rate.
 They obtained
Eq. \ref{PPT_ionization_correct} for the optical cycle-averaged ionization rate $W_{PPT} (E_L, \omega_L)$ for the linear polarized driving laser field
 (from now on and throughout the rest of the paper, unless otherwise explicitly stated, we will use atomic units, $\hbar = e = m_e = 1$).
 
 The refinement of strong field ionization models along with the experimental effort for its validation is an ongoing active field of research \cite{Reiss_effect_PRA_1980, Ammosov_ADK_JETP_1986, Yudin_Ivanov_ion_PRA_2001, Becker_ionization_rates_PRA_2001, Tong_LinCD_Molecular_ADK_PRA2002, Popruzhenko_ionization_arbitrary_laser_PRL_2008, Zhao_intensity_ion_calib_PRA_2016, Larochelle1_Coulomb_ionization_J_Phys_B_1998, Cornaggia_ionization_molecules_PRA_2000, Wu_Dorner_Multiorbital_Tunneling_Ionization_of_the_CO_Molecule_PRL_2012, Lai_SFI_experimentalPRA_2017, Lloyd_Comparison_strong_field_ionization_models_OpExp_2019}.
It has been experimentally shown that ionization rate predicted by the PPT  model is excellent for atoms up to the Keldysh parameter  $\gamma \sim 3-4$ \cite{Larochelle1_Coulomb_ionization_J_Phys_B_1998, Lai_SFI_experimentalPRA_2017} and even valid for small molecules \cite{Cornaggia_ionization_molecules_PRA_2000}.
Another widely used ionization model is the Ammosov, Delone, Krainov (ADK)  that can be obtained from PPT model in the limit of $\gamma\rightarrow 0$. Main advantage of the ADK is its relative simplicity. However, it has been shown experimentally that the validity of the ADK formula is limited to the tunneling regime with Keldish parameter $\gamma \leq 0.5$ \cite{Ilkov_Ionization_atoms_tunnel_J_Phys_B_1992}. 
In this work, we employ the PPT model for calculating the cycle-averaged ionization rate  $W_{01}$ Eq. \ref{PPT_ionization_correct} including the BSI correction Eq. \ref{BSI_ionization_rate}. 

After a neutral atom is ionized by a strong laser, with a probability rate $W_{01}$, resulting in the formation of a singly charged ion, there exists a probability that the remaining ion will undergo further ionization in the laser field, ultimately yielding a doubly charged ion. This phenomenon is referred to as sequential double ionization. The probability rate, denoted as $W_{12}$, can be calculated using the same PPT ionization Eq. \ref{PPT_ionization_correct}, but with the consideration of the ionization potential, $I_p$, for the singly charged ion, which is higher than that of the neutral atom. Additionally, the residual charge is equal to $Z=2$. Consequently, these factors lead to a reduced ionization rate when compared to single-electron ionization.

When an electron gets ejected from the parent ion, there is a probability that the electron wavepacket, driven by the oscillating laser field in the continuum, recollides with the parent ion.
There are four possible outcomes of this process. Firstly, the recolliding electron may experience elastic scattering, or it may contribute to the HHG and generation of attosecond pulses  \cite{Corkum_3step_PRL1993}. 
This is what makes our approach particularly useful in attosecond science, as it enables accurate in situ characterization of the spatially resolved intensity and duration of strong-field femtosecond pulses within the interaction volume.
However, these processes do not contribute to ionization in the medium after the laser pulse is gone.

The recolliding electron wavepacket may also cause either the ionization of the parent ion, which is referenced as non-sequential double ionization (NSDI), with a rate of $W_{NSDI}(E(t))$ or ion's excitation to a higher energy state, denoted as $1^*$, with a rate $W_{NSE}(E(t))$, and its  subsequent ionization (RESI) \cite{Kopold_RESI_PRL_2000, Feuerstein_RESI_PRL_2001, Bergues_NSDI_NatComm_2012}, with a rate $W_{1^* 2}(E(t))$ . This $W_{1^* 2}$ ionization rate is calculated again using the PPT ionization model with the ionization potential of singly charged ion in the excited state.

To account for NSDI and the RESI processes in ionization dynamics, we implement a semiclassical model for the calculations of the $W_{NSDI}(E(t))$ and $W_{NSE}(E(t))$ rates \cite{Bergues_NSDI_NatComm_2012}. In the nutshell, we calculate the probability of ionization $P(t_i)$ in the time window $[t_i \; t_i+dt]$ within an optical cycle of period T, $P(t_i)=W(E(t_i))\bar{P}(0\leq t\leq t_i) dt = W(E(t_i)) \exp{[-\int_0^{t_i} W(E(t'))dt']} dt $, where $W(E(t_i))$ is an instantaneous ionization rate Eq. \ref{inst_ionization_rate} and  $\bar{P}(0\leq t\leq t_i)=\exp{[-\int_0^{t_i} W(E(t'))dt']}$ is the probability that no ionization occurres in the $[0 \; t_i]$ time window.   For each ionization time, we calculate the classical trajectory of electron within an oscillating field $E(t)=A(t) \cos{(\omega t)}$, the return time $t_r$ and return energy $E_k(t_r)$. $A(t)$ is a slowly varying envelope of the pulse that is assumed to be constant within a time scale of one optical period (the so-called slowly varying envelope approximation (SVEA) \cite{Diels_Ultrafast_2006}). In semiclassical picture, upon ionization the electron wavepacket emerges from the tunnel barrier with a momentum coordinates, $p_{\perp}$, perpendicular to classical trajectory direction (in the atomic units) \cite{DELONE_ionization_atoms_JOSA1991, Ivanov_Anatomy_J_Mod_Opt2005}:
\begin{equation}
\label{p_perp}
\psi(p_{\perp}) = \psi(0) \exp {\left(-\frac{p_{\perp}^2 T_{\textrm{tunnel}}}{2}\right)}.
\end{equation}
This wavepacket spreads as it propagates in the continuum during the $t_r-t_i$ time window. Upon recollision, we calculate the overlap of the recolliding wavepacket with the respective energy $E_k(t_r)$ - dependent cross-sections: $\sigma_{NSDI}\left(E_k(t_r)\right)$, the cross-section of electron impact ionization of ion, and $\sigma_{NSE}\left(E_k(t_r)\right)$, the cross-section of electron impact excitation of ion \cite{Tawara_X_section_1987, Jesus_Atomic_JOSA_B_2004}.  We determine the probability of the process of interest by examining the overlap, as derived in the section "Non-sequential double ionization" of the Methods. The rates $W_{NSDI}$ and $W_{NSE}$  are found by normalizing the respective probabilities by half the period of the driving laser field.

When laser field approaches or exceeds the critical value of $E_{BSI}=I_p^2/(4Z)$, the ionization potential $I_p$ reaches or surpasses the potential barrier created by the combination of Coulomb attraction to the atomic core with charge Z and the instantaneous laser field.
In this so-called barrier-suppression ionization (BSI) region, it is well-known that strong-field ionization models  \cite{Perelomov_PPT3_JETP1967, Ammosov_ADK_JETP_1986} overestimate the ionization rate \cite{Ilkov_Ionization_atoms_tunnel_J_Phys_B_1992, Delone_tunneling_Physics_Uspekhi_1998}. To account for this discrepancy, we use a correction factor \cite{Zhang_Empirical_formula_BSI_PRA_2014} for ionization rate calculations Eq. \ref{BSI_ionization_rate}, as detailed in the section "Barrier-suppression ionization correction" of the Methods. 

Taking into account all the aforementioned processes, we derive the following set of coupled equations that describe the ionization dynamics:


%

\begin{empheq}[left=\empheqlbrace]{align}
\frac{dN_0(\vec{r},t)}{dt} &= -W_{01}(E(\vec{r},t))N_0(t)-W_{NSDI}(E(\vec{r},t))N_0(\vec{r},t)-W_{NSE}(E(\vec{r},t))N_0(\vec{r},t), \nonumber \\
\frac{dN_1(\vec{r},t)}{dt} &=  W_{01}(E(\vec{r},t))N_0(t) - W_{12}(E(\vec{r},t))N_1(\vec{r},t), \nonumber \\
\frac{dN_{1^*}(\vec{r},t)}{dt} &= W_{NSE}(E(\vec{r},t))N_0(\vec{r},t)-W_{1^* 2}(E(\vec{r},t))N_{1^*}(\vec{r},t), \nonumber \\
\frac{dN_2(\vec{r},t)}{dt} &= W_{NSDI}(E(\vec{r},t))N_0(t)+W_{12}(E(\vec{r},t))N_1(t)+W_{1^* 2}(\vec{r},t)N_{1^*}(\vec{r},t),
\label{Rate_equations}  
\end{empheq}
where $N_0(\vec{r},t)$, $N_1(\vec{r},t)$, $N_{1^*}(\vec{r},t)$, and $N_2(\vec{r},t)$ are the densities of neutral, singly ionized, singly ionized in the excited state and doubly ionized atoms/molecules, respectively.

 \begin{figure}
    \begin{center}
	\includegraphics[width=1\linewidth]{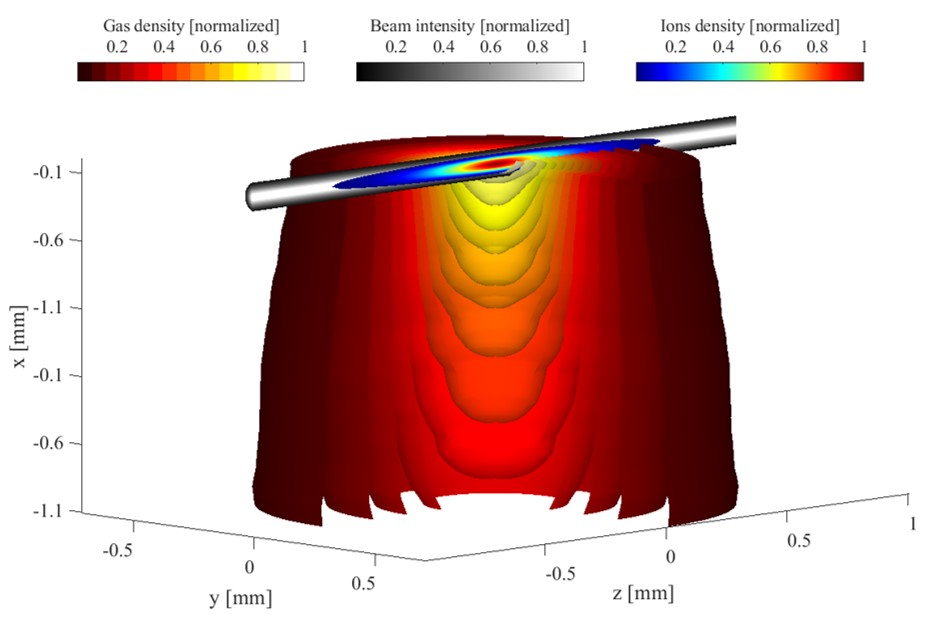}
    \caption{\label{Figure_BeamIons_inJet} Ion charge density overlaid with the beam intensity distribution in the interaction volume, shown within the measured and tomographically reconstructed normalized argon gas density in the pulsed jet (back pressure 7.5 bar, laser pulse energy 0.5 mJ). The maximum ions charge density is $\rho_{ions}^\text{MAX}= 3\times 10^{15}\text{cm}^{-3}$, and the maximum gas density is $N_0^\text{MAX}=8.6 \times 10^{15}\text{cm}^{-3}$.  Gas jet is visualized as cut-outs within isosurfaces of constant gas density.
  }
    \end{center}
\end{figure}

To comprehensively elucidate the ionization dynamics within the focal volume in both space and time, we solve the system of Eqs. (\ref{Rate_equations}) taking into account the measured intensity cross-section of the laser beam in the focus as shown in Fig. \ref{Fig_BeamFocusProfile}, as well as the change of gas jet density along the laser propagation direction within the jet. The gas density distribution, $N_0(\vec{r},t=0)$ within the jet has been measured using the laser induced strong-field ionization gas jet tomography \cite{Tchulov_LaserStrongFieldTom_ScRep_2017} by scanning the laser beam across the slice of interest and measuring the ion signal, as depicted in Fig. \ref{Figure_BeamIons_inJet}. 
The spatial intensity distribution of the laser beam couples into the ionization dynamics equations via spatial dependence of respective ionization rates, $W(E(\vec{r},t))$. 
 
 \begin{figure}
    \begin{center}
	\includegraphics[width=1\linewidth]{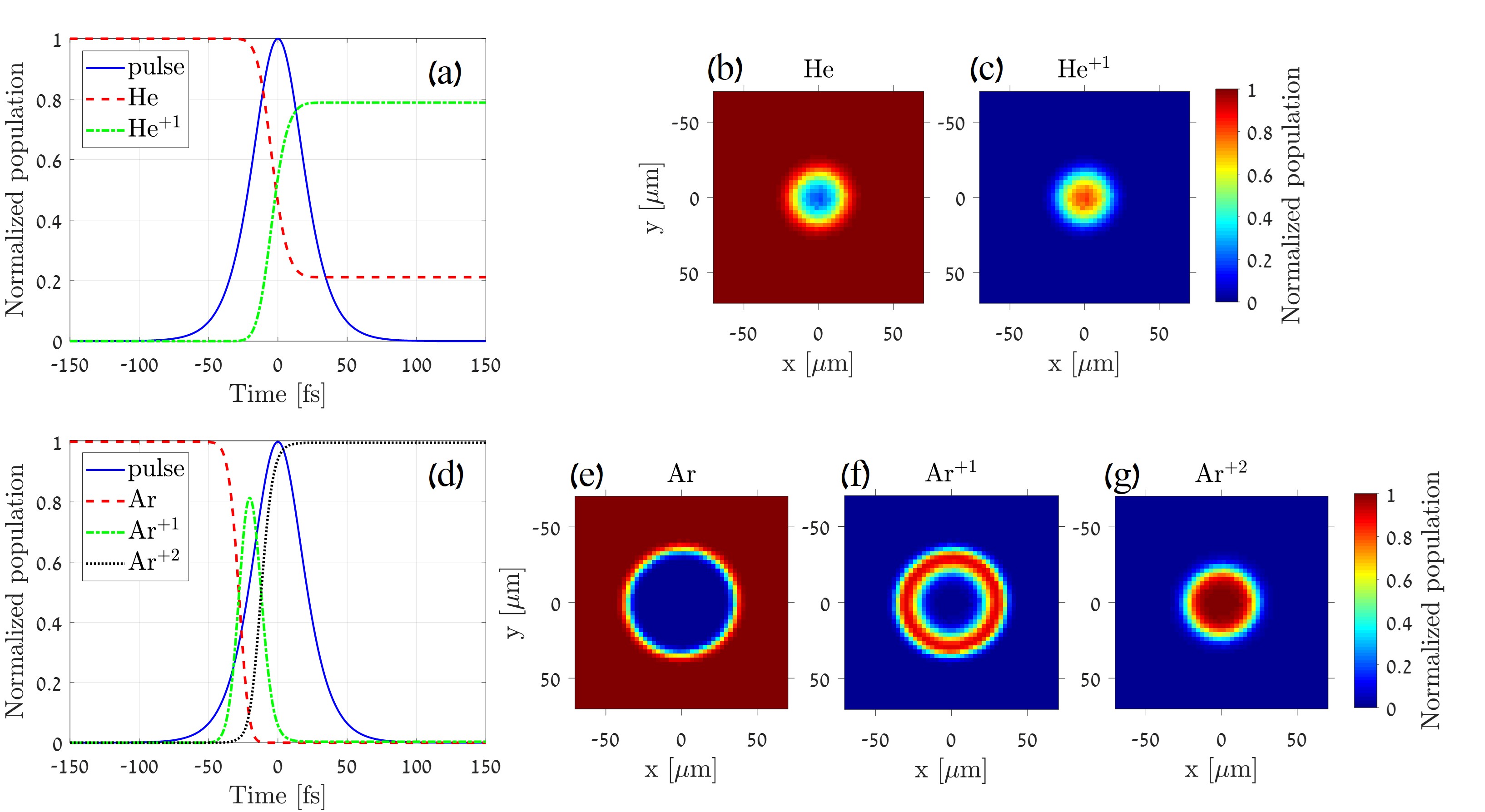}
    \caption{\label{Figure_space_time_ionization}   Ionization dynamics in space and time in the focal volume for argon (a,b,c) and helium (d,e,f,g). The temporal evolution of the spatial ion distributions is visualized in the Supplementary movie [?file-link?]. This movie provides further insights into the dynamics observed in the figure. 
  }
    \end{center}
\end{figure}
 
Figure \ref{Figure_space_time_ionization} illustrates the typical ionization dynamics for helium and argon at the focus of a femtosecond laser pulse.
For this analysis, we assume a typical for a femtosecond laser  $f^2(t)=\textrm{sech}^2(t/\tau)$ temporal intensity pulse profile \cite{Zhou_Amplifier_sech2_OL_1995} ,  with a FWHM  pulse duration of 43 fs and pulse energy of 2 mJ. The rationale for choosing these pulse parameters becomes evident from the discussion below. For pulse durations in this range, we can safely assume space-time invariance \cite{Bor_Distortion_Femto_OptCom_1992, Porras_charact_E_field_OL_2009, Attia_space_time_CEP_OE_2022}, i.e. $E(\vec{r}_\bot,t)=E_{peak} \varepsilon(\vec{r}_\bot) f(t)$, where $E_{peak}$ is the electric field peak amplitude, $\varepsilon(\vec{r}_\bot)= \sqrt{I_N(\vec{r}_\bot)}$, with $I_N(\vec{r}_\bot)$ being the measured (as depicted in Fig. \ref{Fig_BeamFocusProfile}) normalized focused beam intensity distribution.

Figure \ref{Figure_space_time_ionization} (a) depicts the time evolution  of the laser intensity, the density of the neutral helium atoms, and the density of the singly ionized He atoms at the center of the driving laser beam.
Due to the very high ionization potential of helium, double ionization is negligible for the considered laser pulse, and ionization of neutral atoms with the $W_{01}$ rate is the dominant process. Consequently, for helium, the sum of neutrals and singly ionized atoms' density remains constant. It is evident that the ionization becomes significant before the peak of the laser pulse.
In this example, approximately $20\%$ of the initial helium gas gets ionized, $~10$ fs before the pulse reaches its peak intensity (at this point in time, the intensity reaches $~85\%$ of its peak). 
After the laser pulse has passed, approximately  $80\%$ of the gas at the beam's center has been ionized. This creates a distinct 'hole-burning' effect in the spatial distribution of neutral atoms, as shown in Fig. \ref{Figure_space_time_ionization} (b), complemented by the generation of spatially "molehill"-like localized  singly ionized helium atoms, whose normalized density is visualized in Fig. \ref{Figure_space_time_ionization} (c).

Figure \ref{Figure_space_time_ionization} (d) presents the time evolution of the ionization process for argon at the beam's center, driven by the same laser pulse used in the helium example.
Notably, while helium reached $70\%$ ionization only past the peak intensity of the laser pulse on its falling slope, argon achieves this same level by the time the femtosecond pulse reaches just $40\%$ of its peak intensity on the leading edge of the pulse.
 This significant difference clearly arises from argon's lower ionization potential compared to helium. 
 
 These observations underscores the fact that strong-field ionization exhibits high sensitivity not only to the peak laser intensity but also to the temporal evolution of the driving pulse, 
 particularly within the time window around the peak where the majority of the gas undergoes ionization.

Another major consequence of argon's lower ionization potential compared to helium is that double ionization plays a significant roles in the ionization dynamics of argon.
This is in striking contrast to helium, where the impact of these processes on ionization was negligibly small.
Initially, as the laser pulse intensity increases, strong-field single ionization begins to dominate the ionization dynamics.
This phenomenon leads to an increase in the density ($N_1(t)$) of singly ionized $Ar^{+1}$ ions, naturally accompanied by a decrease in the density of neutral atoms ($N_0(t)$), as illustrated in Fig. \ref{Figure_space_time_ionization} (d).
With the increasing density of $Ar^{+1}$ ions and laser intensity, double ionization  gain prominence, leading to the double ionization of $Ar^{+1}$ ions (depicted as the dotted black curve labeled $Ar^{+2}$ in the figure).
 
This phenomenon depletes the transient population of singly ionized $Ar^{+1}$ ions at the beam's center. 
Notably, the duration of the transient $Ar^{+1}$ population at the beam's center is only 19 fs (FWHM), significantly shorter than the 43 fs laser pulse duration. 
While the high laser intensity at the center drives double ionization, it weakens with increasing distance further away from the center, leading to single ionization there as the dominant process.

This spatial dependence results in the characteristic "molerun"-like distribution of  $Ar^{+1}$ ion density across the focal plane once the laser pulse has passed, as illustrated in Figure \ref{Figure_space_time_ionization} (f).
This phenomenon is accompanied by a localized "molehill" peak of $Ar^{+2}$ ions (Fig. \ref{Figure_space_time_ionization} (g)) and a significantly more depleted "hole-burning" compared to helium (Fig. \ref{Figure_space_time_ionization} (e)).

\begin{figure}
    \begin{center}
	\includegraphics[width=1\linewidth]{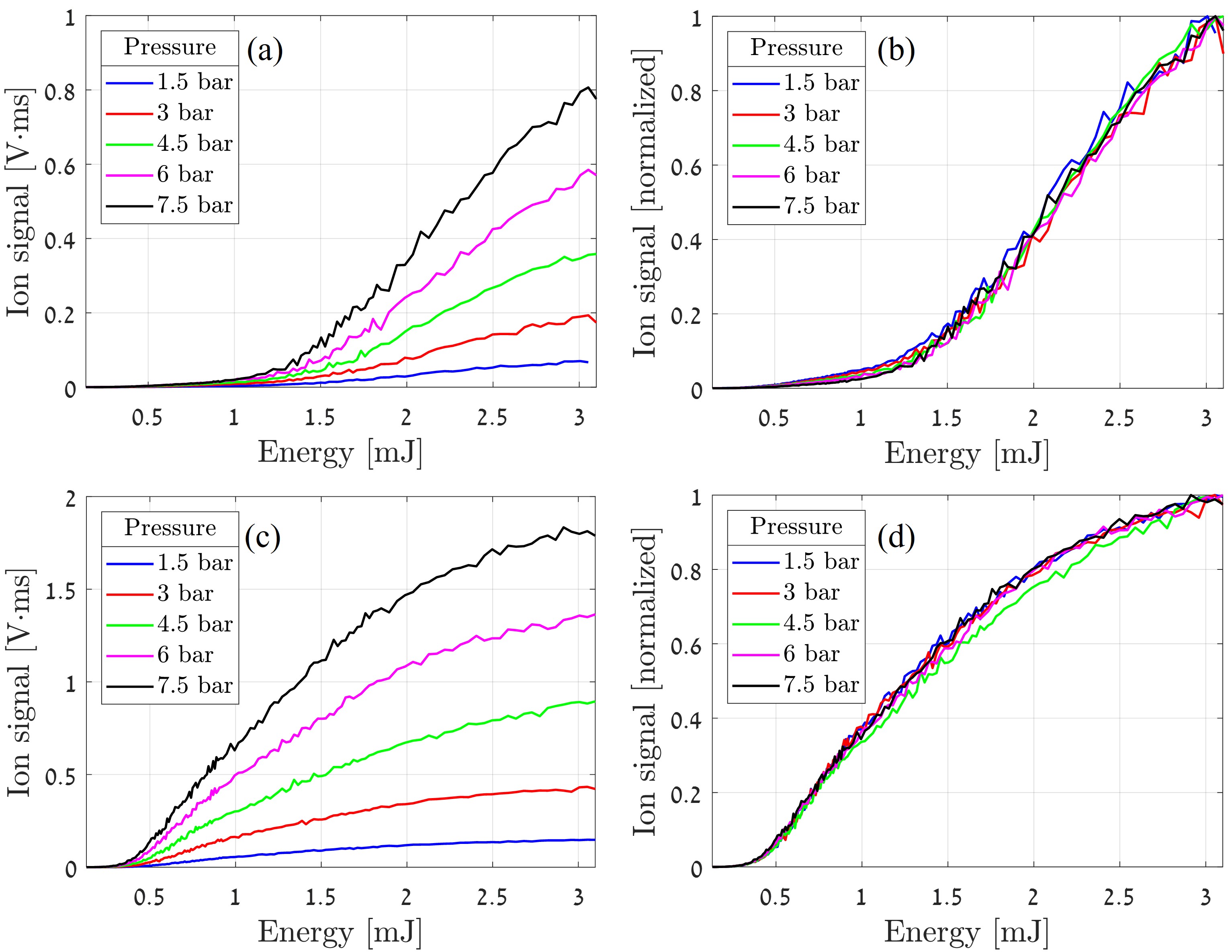}
    \caption{\label{Figure_IonSignal}   Measured ion signal vs. laser pulse energy. (a) In helium (b) In helium normalized (c) In argon (d) In argon normalized.
  }
    \end{center}
\end{figure}


\subsection{Experimental Results}
 We now turn to the experimental results of strong field ionization measurements as a function of laser pulse energy. Figures \ref{Figure_IonSignal} (a) and (c) show the measured ion signal for helium and argon, respectively, at various back pressures in the pulsed valve, ranging from 1.5 bar to 7.5 bar. As expected, increasing back pressure leads to a higher gas density in the focal volume of the gas jet, consequently enhancing the ion signal for a given gas. Additionally, at the same pulse energy, the argon signal surpasses that of helium due to the difference in their ionization potentials.

 To gain deeper insights, we normalized the same measurements as shown in Figs. \ref{Figure_IonSignal}(b) and (d).  Surprisingly, our experiment reveals that the normalized ionization curves for both helium and argon exhibit no discernible dependence on the gas density within the jet. Notably, apart from a single outlier point at 4.5 bar for argon, all ionization curves remain strikingly close within the expected range of random measurement error.
 We attribute this lone outlier among the ten measurements in both helium and argon to an anomaly in the driving laser stability and exclude it from further consideration.
 
 It's important to note that after the laser pulse is over, the multitude of recombination processes may occur within the ionized plasma, such as three-body recombination, radiative recombination,  and others \cite{Raizer_Gas_Discharge_Physics_Book_1991}.
Accompanied by space charge effects, such as Debye  shielding \cite{Fitzpatrick_Plasma_Physics_2022} and ambipolar diffusion \cite{Raizer_Gas_Discharge_Physics_Book_1991}, which disrupts the charge current and separation of ions and electrons within the laser-induced plasma subjected to the electric field of an ion detector, these recombination processes can decrease the fraction of ions and electrons reaching the detector plates and consequently diminish the measured ion signal. 

One might anticipate a pronounced, generally nonlinear influence of gas density on the measured ion signal due to the interplay of these processes, potentially altering the ionization curve shapes. However, our experimental results clearly indicate that it is not the case---the normalized ionization curves exhibit remarkable resilience and remain independent of gas density in the jet within the studied range.

 
 Considering that the ion signal is proportional to the total ionized charge as illustrated in Fig.  \ref{Fig_ion_signal} (b), we calculate and fit the theoretically calculated ion yield, $Y_{ions}$, to the measured normalized ion signals shown in Figs. \ref{Figure_IonSignal} (b) and (d) for  every back pressure for both helium and argon.
 The ion yield is determined by the integral of ion charge  density, $\rho_{ions}(\vec{r})$ (as depicted in Fig. \ref{Figure_BeamIons_inJet}), over the focal volume: $Y_{ions} = \iiint_{V_{focal}} \rho_{ions}(\vec{r})dV$. The ion charge density, after the ionizing laser pulse has passed, is given by
 $\rho_{ions}(\vec{r})=\left(N_1(\vec{r}\right)+N_{1^*}(\vec{r}))+2N_2(\vec{r})$, where $N_1(\vec{r})$, $N_{1^*}(\vec{r})$ and $N_2(\vec{r})$ represent singly and doubly ionized densities  calculated at the end of the pulse by solving the set of Eqs. \ref{Rate_equations}, as has been described. 
 
 Note that we measure both the total energy of the pulse as well as spatial intensity distribution in the focal volume, as shown in Fig. \ref{Fig_BeamFocusProfile}. For the measured charge and gas density distribution in the generated plasma, our analysis shows that  the spatial distortion of the beam in the interaction region is negligible. Therefore, for a specific pulse shape, the only free parameter in the fitting procedure is the pulse duration, $\tau_{FWHM}$, or equivalently, the peak intensity $I_{peak}$. More details about the fitting procedure are provided in the Methods. 

\begin{figure}
    \begin{center}
	\includegraphics[width=1\linewidth]{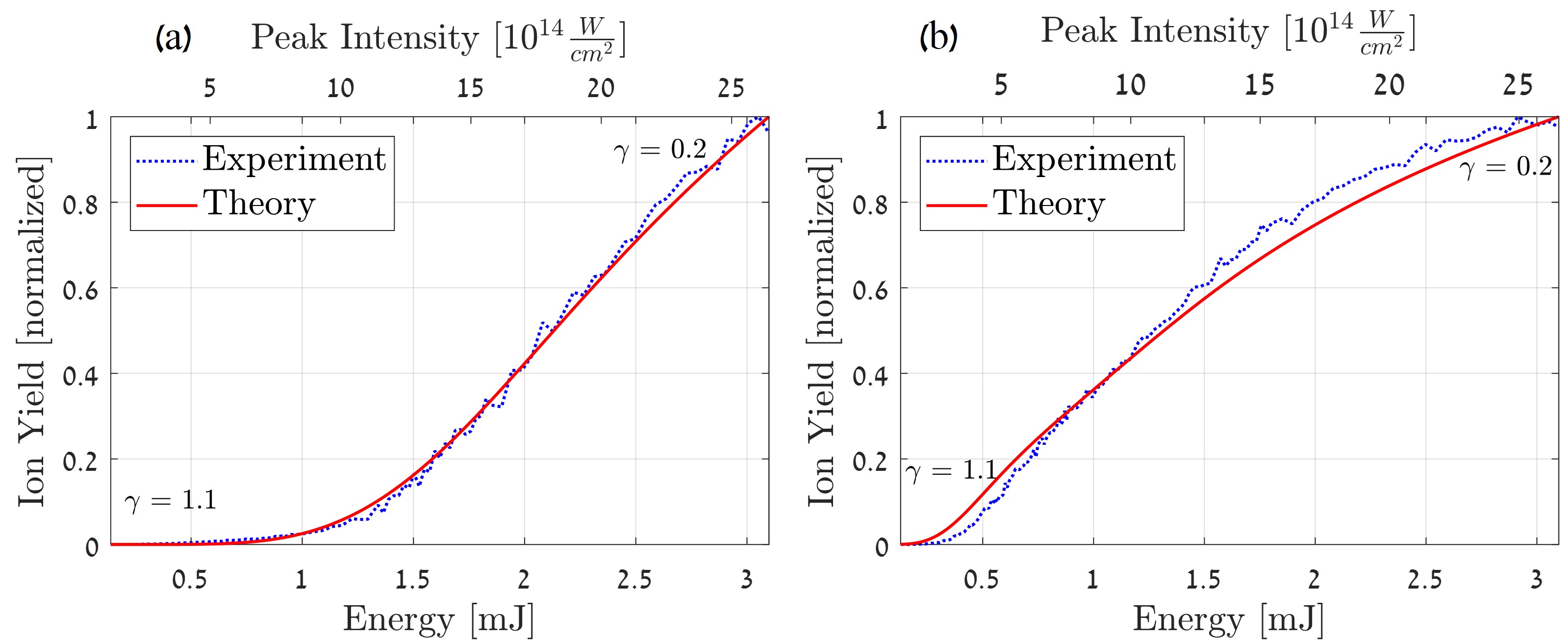}
    \caption{\label{Fig_ion_yield_7_5barr} The strong field ionization model fitted to the measured ionization curves at 7.5 bar back-pressure in (a) helium and (b) argon. For helium, the extracted pulse duration is $\tau_{FWHM} = (41.7\pm 0.5)$ fs with a goodness of fit $R^2 = 0.997$. For argon the extracted pulse duration is $\tau_{FWHM} = (43.3 \pm 1.6)$ fs with a goodness of fit $R^2 = 0.989$. Both intensity and $\gamma$ were calculated using $\tau_{FWHM}$.}
    \end{center}
\end{figure}

 Figure \ref{Fig_ion_yield_7_5barr} (a) shows a typical result of this fitting procedure in helium at a back-pressure of 7.5 bar. The optimal fit yields a full width at half maximum pulse duration of $\tau_{FWHM} = (41.7\pm 0.5)$ fs. The upper horizontal axis in the figure displays the pulse peak intensity corresponding to this pulse duration.
  
 It's noteworthy that increasing the ionizing pulse energy generates a plasma with higher density and larger volume within the gas jet by the time the driving laser pulse ends. This, in turn, can potentially amplify recombination and space-charge effects, as previously discussed, and subsequently diminish the measured ion signal at higher intensities. In the presented version, our theoretical model does not incorporate these macroscopic recombination or space-charge phenomena. Therefore, one might anticipate a systematic error that increases at higher intensities, potentially distorting the shape of the measured signal and causing a deviation from the theoretically calculated curve.
 
 However, our results unequivocally demonstrate the opposite---the normalized measured ionization curves exhibit remarkable agreement with our theoretical model, characterized by a parametric fit error of $\pm 0.5$ fs ($95\%$ confidence interval)  and  goodness of fit of $R^2=0.997$, even at high pulse energies and gas back-pressure.
 In principle, this pulse duration uncertainty encompass both statistical errors and systematic errors. By repeating the measurements, we find a statistical errors to be within $~1$ fs range for single-pulse energy scan across 150 points spanning the energy range.
To isolate systematic errors arising from theoretical model imperfections,  we analyzed and presented in Figs. \ref{Fig_ion_yield_7_5barr} data averaged over 100 ion signal measurements for each pulse energy, effectively reducing statistical error to 0.1 fs.

 Only at low pulse energies ($<~0.8$ mJ) and correspondingly very low ionization yield do we measure ionization yield in helium higher than predicted by the theory. This is attributed to the ionization of the residual gas impurities in helium, though not visible on the scale of the plot and having negligible effect on the fitting, as these parasitic contributions are very low compared to the dynamic range of the measured ion signal. Given the high ionization potential of helium, at low intensities, residual gases such as nitrogen, oxygen, and water vapor mainly contribute to the ionization signal, surpassing the minor contribution from ionization of helium itself. 

 Figure \ref{Fig_ion_yield_7_5barr}(b) depicts a representative result of the analogous fitting procedure in argon at the same back-pressure of 7.5 bar. The best fit results in a pulse duration of  $\tau_{FWHM} = (43.3\pm 1.6)$ fs.
  In contrast to helium, where single-electron ionization dominates across the measured intensity range due to its high ionization potential, double ionization phenomena comes into play in argon. Our calculations show that SDI plays a major role, followed by  NSDI, while RESI contribution was negligible in the relevant intensity range. Additionally, BSI has been found to play significant role as well.
 This complex interplay also explains why, while the fitting results for argon remain good with a pulse duration fit error of $\pm 1.6$ fs and $R^2 = 0.989$, they fall slightly short of the exceptional goodness of fit achieved for helium.
  Remarkably, despite this intricate interplay of ionization processes in the model, we obtained a pulse duration in argon that closely matches that measured in helium.
   
   It is crucial to include these strong field ionization phenomena in the computations; neglecting them leads to an inconsistent pulse duration in argon compared to the helium result, indicating a deficiency in the comprehensive ionization model.  For instance, not accounting for sequential double ionization in argon results in a pulse duration of 74 fs, a stark $>70\%$ deviation from the helium ($~42$ fs) or well-modeled argon ($~43$ fs) result. Omitting BSI only in argon leads to 58 fs, still very significant  38\% deviation.
   
  At the same time, we validated that neglecting SDI (or any double ionization process) in helium ionization calculations made, as expected, no difference to the pulse duration result.
  Neglecting NSDI doubles the systematic error in argon compared to helium, but this effect is significantly less important than the SDI contribution. This is because the contribution of  $W_{12}$ dominates over  $W_{NSDI}$ in argon already at the peak intensity of $5 \times 10^{14}$ $\textrm{W}/\textrm{cm}^2$.

\begin{figure}
    \begin{center}
	\includegraphics[width=0.6\linewidth]{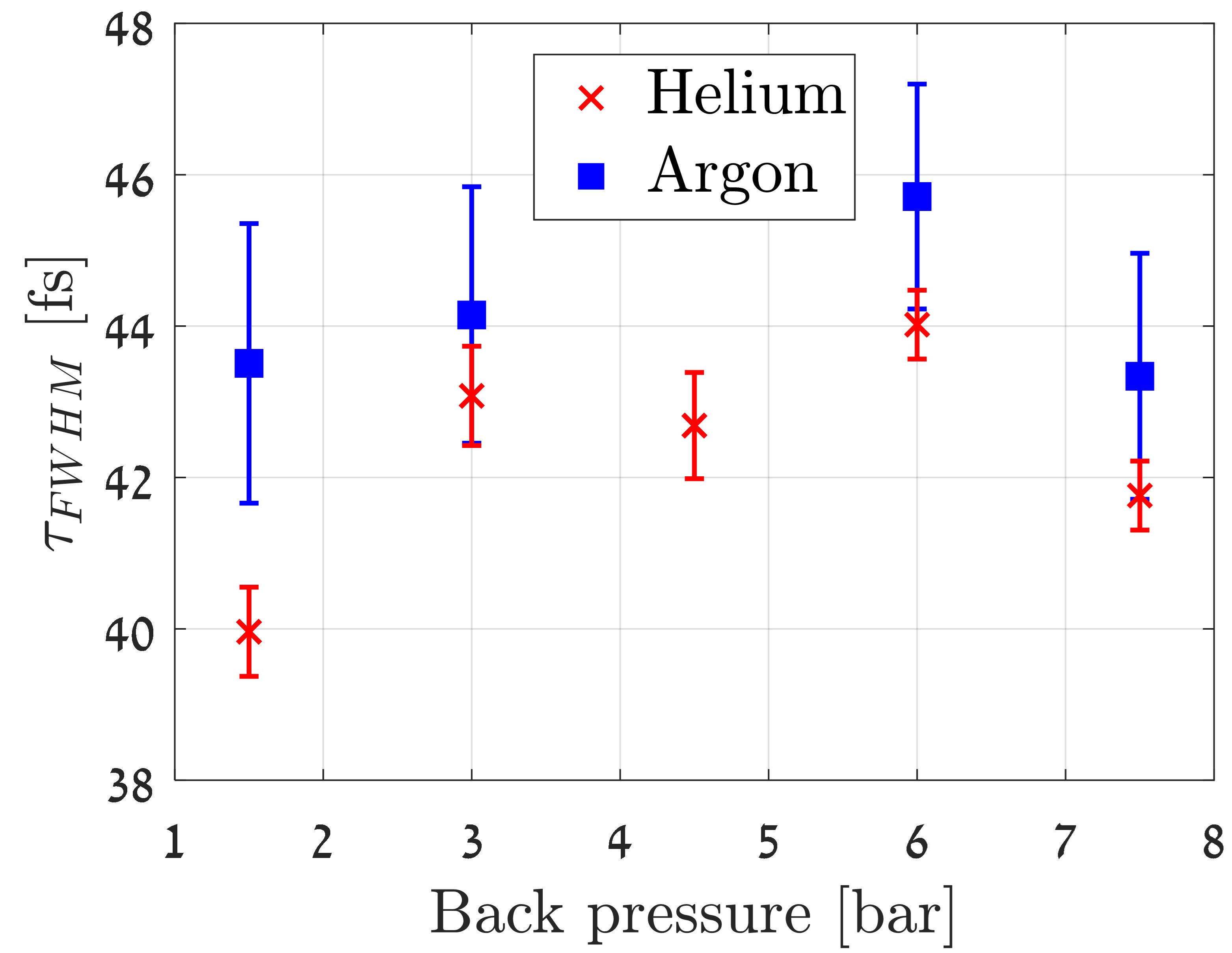}
    \caption{\label{Fig_tau_vs_pressure} 
   Pulse duration $\tau_{FWHM}$ extracted by fitting the theoretical model to the measured ionization curves for helium (red crosses) and argon (blue squares). The error bars indicate a parametric fit error ($95\%$ confidence interval).   
  }
    \end{center}
\end{figure}

Figure \ref{Fig_tau_vs_pressure} summarizes the pulse durations extracted for both helium and argon across various back pressures. 
Notably, the average pulse duration across all gases and pressures is  $\tau_{FWHM}=43.1\pm 1.6$ fs. 

As each data point in the plot is shown after averaging reducing statistical error to 0.1 fs, the error bars indicate the pulse duration parametric fit error.    
The average value for pulse duration for the measurements in He is $42.3\pm 1.5$ fs and in argon $44\pm 1$ fs. 
Thus we find that each of the systematic errors due to imperfections in theoretical model, the variations in the back pressure and usage of different gases does not exceed 2 fs or is less than $5\%$ of the measured pulse duration.
The main source  of statistical error in our setup was caused by beam pointing stability issues---this can be further improved, for instance, by implementing active beam stabilization techniques \cite{Grafstrom_beam_pos_control_OptCom1988}.
 However, for most practical purposes, the demonstrated statistical error (approximately 1 fs), even without averaging, is already satisfactory and comparable to the systematic errors.

\section{Discussion}

The pulse duration at the output of our regenerative amplifier is 30 fs; however, in situ measurements using our scheme reveal a pulse length  $~40\%$ longer. Since peak intensity is inversely proportional to pulse duration,  relying solely on the output pulse duration would result in a significant error  in estimating the peak intensity within the interaction region. This once again underscores the importance of in situ measurements for both pulse duration and peak intensity, especially in strong field and attosecond science experiments.



To assess the impact of temporal pulse shape on our results, we considered both the typical $\textrm{sech}^2$  and Gaussian profiles used to characterize femtosecond laser outputs.
Additionally, we modeled the influence of strong dispersion-induced pulse distortions, by propagating 30 fs pulse through 50 mm fused silica.
 Notably, the estimated laser peak intensity remained remarkably robust, showing a maximum variation of only 3\% across different pulse shapes and distortions.

It is important to stress that the measured ion signal depends not only on the peak intensity, but also on the pulse duration.
As the strong field ionization is highly nonlinear process, the shape around the peak of the pulse becomes particularly important.  
Interestingly, for $\textrm{sech}^2$, Gaussian, and dispersion-distorted pulses with the same FWHM, the pulse intensity varies by no more than $2\%$ of its peak within the upper half of the intensity profile, where most ion signal is generated. 
However, the main variation in intensity between these pulses occurs around their off-peak regions, where intensity is low and impact on the strong field ionization is minor.
This further underscores an important advantage of our technique for strong-field ionization experiments: it is inherently sensitive to the pulse characteristic most crucial for the process, namely the intensity profile around the peak, where the majority of strong field ionization phenomena take place.

The robustness of our method to strong dispersion distortions is crucial for practical applications, as the pulse's spectral phase can be significantly affected by various factors. These include, among others, dispersion from optics and propagation in the air, suboptimal compressor grating settings, and potential distortions from target gases. While the target gas distortions were not significant in the specific experiment being discussed, the method's resilience to such factors remains important for ensuring accurate and reliable results in diverse real-world scenarios. By validating this robustness, the method can be more effectively applied across different systems and environments, enhancing its overall usefulness.

Utilizing the presented ionization model, our method is valid within the slowly varying envelope approximation (SVEA) \cite{Diels_Ultrafast_2006}, i.e. $ \lvert dA(t)/dt \rvert \ll \omega_L \lvert A(t) \rvert$, indicating that the variation of the electric field envelope within an optical cycle is negligible. 
However, it is important to stress that satisfying the SVEA condition is crucial only within a time window around the peak of the laser pulse, where the ionization rate, and therefore the electric field, are substantial, thereby relaxing the SVEA requirement.
For example, in our experiments using a laser with a central wavelength of 800 nm, our analysis shows that these criteria are well satisfied for pulses with durations $\geq 10$ fs.
One interesting direction for future research involves refining the ionization model to expand its validity to the near-single-cycle limit of femtosecond laser pulses, with controlled carrier envelope phase (CEP) \cite{Jones_CEP_modelocking_Science_2000}, and taking into account the possible impact of space-time coupling \cite{Attia_space_time_CEP_OE_2022}.




In conclusion, the accuracy of strong-field experiments, particularly in attosecond science, should no longer be constrained  by uncertainties in the femtosecond laser beam characteristics within the interaction region.
Spatially resolved, in situ intensity measurements of the driving laser, combined with the measurement of spectrally resolved HHG wavefronts \cite{Frumker_sword_OL2009}, pave the way for resolving a long-standing issue in attosecond science experiments: the spatial averaged measurements across the focal volume  \cite{Frumker_wavefronts_OE2012}.
This approach holds the potential  to significantly enhance experimental precision, for instance, in imaging orbitals \cite{Itatani_tomog_Nature2004, Vozzi_gen_tomography_NatPhys_2011}, tracing molecular dynamics \cite{Li_dynamics_N2O4_Science_2008}, identifying and time resolving tunneling wave packets \cite{Smirnova_multielectron_Nature2009}, following Auger decay \cite{Drescher_Auger_Nature_2002}, and probing polar molecules with high harmonic spectroscopy \cite{Frumker_orientation_HHG_PRL2012}---enabling access to a single-molecule response in a strong laser field \cite{Frumker_wavefronts_OE2012}.

Helium's exceptionally high $I_p$ allows the single-electron PPT model to adequately describe its ionization within the relevant intensity range. However, most other molecules and even noble gas atoms possess significantly lower $I_p$---similar or even lower than argon.  To comprehensively elucidate the ionization dynamics of such systems,  accounting for double ionization and BSI becomes crucial. Frequently, ion measurements have been interpreted solely through single-electron strong-field ionization models, considering only peak intensity ionization rates; our findings clearly illustrate the inadequacy of this approach. 


Furthermore, the implications of our findings extend beyond improving the fidelity of strong-field and attosecond science experiments. 
The observed robustness to pressure variations and remarkable agreement between our measurements and the single-atom ionization model for helium suggest that ion detection approach can serve as a valuable experimental tool for validating and refining theoretical models of strong-field ionization for more complex atoms and molecules.

\section{Materials and Methods}
\subsection{Strong-field ionization rate PPT}

In this section, we explicitly summarize the formulas for the optical cycled averaged ionization rate, employing the Perelomov, Popov, Terentev (PPT) formulas \cite{Perelomov_PPT_JETP1966}, which are used in this work with certain corrections:

%
%
%
%
 
\begin{multline}
\label{PPT_ionization_correct}
W_{mPPT}(E_L, \omega_L) = C_{n^* l^*}^2 f_{lm} I_p \sqrt{\frac{6}{\pi}} \left( \frac{2E_0}{E_L} \right)^{2n^*-|m|-3/2} \\
\times (1+ \gamma^2)^{|m|/2+3/4} A_m (\omega_L, \gamma) \\
\times \exp{\left( - \frac{2E_0}{3 E_L} g (\gamma)  \right)},
\end{multline}

where:

$E_L$ is the laser field,

$\omega_L$ is the laser cycle frequency,

$n^*$ is the effective principal quantum number:
\begin{equation}
\label{n_star_factor}
 n^* = \frac{Z}{\sqrt{2 I_p}},
 \end{equation}

$l^*$ is the effective orbital quantum number:
\begin{equation}
\label{l_star_factor}
 l^* =n^* - 1 ,
 \end{equation}

%

$E_0$ is the characteristic atomic electric field:
\begin{equation}
\label{E_0}
 E_0 =\left(2I_p \right)^\frac{3}{2} ,
 \end{equation}

\begin{equation}
\label{c_2_factor}
 C_{n^* l^*}^2 = \frac{2^{2n^*}}{n^* \Gamma (n^*+l^*+1) \Gamma (n^* -l^*)},
 \end{equation}

 \begin{equation}
\label{f_lm_factor}
 f_{lm} = \frac{(2l+1)(l+|m|)!}{2^{|m|} |m|! (l-|m|)!},
 \end{equation}

\begin{equation}
\label{A_factor}
A_m (\omega_L, \gamma)= \frac{4 }{\sqrt{3 \pi}|m|!} \frac{\gamma ^2}{1+ \gamma ^2} \sum_{k \geq k_{min}}^{+\infty}  e^{-\alpha (\gamma) (k- k_{min})} \xi _m \left( \sqrt{\beta (\gamma) (k- k_{min})} \right)
\end{equation}

\begin{equation}
\label{alpha_factor}
 \alpha (\gamma) =  2 \left[ \textrm{arcsinh} ( \gamma) - \frac{\gamma}{\sqrt{1 + \gamma ^2}} \right] ,
 \end{equation}

\begin{equation}
\label{beta_factor}
 \beta (\gamma) = \frac{2 \gamma}{\sqrt{1 + \gamma ^2}} ,
 \end{equation}

\begin{equation}
\label{xi_factor}
 \xi _m (x) = \frac{x ^{2|m|+1}}{2} \int^1_0 \frac{e^{-x^2 t} t^{|m|}}{\sqrt{1-t}} \, dt ,
 \end{equation}

\begin{equation}
\label{g_factor}
 g(\gamma) = \frac{3}{2 \gamma} \left[\left(1+ \frac{1}{2 \gamma ^2} \right) \textrm{arcsinh} ( \gamma) - \frac{\sqrt{1 + \gamma ^2}}{2 \gamma} \right],
 \end{equation}
where k  is the number of above threshold ionization photons (ATI) \cite{NIST_ASD_database_2017}, starting with the minimum number of ATI photons, $k_{min}$ (this formula for clearity is written in the SI units):
\begin{equation}
\label{k_min_photons}
k_{min} = \lceil{\frac{I_p + U_p}{\hbar \omega_L}}\rceil=\lceil{\frac{I_p}{\hbar \omega}\left({1+\frac{1}{2 \gamma^2}}\right)}\rceil; U_p= \frac{e^2 E_L^2}{4 m_e \omega_L^2},
\end{equation}
Here, $U_p$ is a ponderomotive energy, which is the cycle-averaged kinetic energy of the electron motion in the field. The $k_{min}$  is rounded up to the nearest integer here, whereas in the original PPT paper \cite{Perelomov_PPT_JETP1966}, $k_{min}$  was not rounded, which leads to unphysical results.

The simplified cycled averaged ADK formula for the ionization rate, $W_{ADK}(E_L, \omega_L)$, is often used \cite{Ammosov_ADK_JETP_1986}  (can be obtained from $W_{PPT}(E_L, \omega_L)$ in the limit $\gamma \rightarrow 0$ as shown in \cite{Perelomov_PPT_JETP1966} Eq. 59):

\begin{equation}
\label{ADK_ionization_rate}
W_{mADK}(E_L, \omega_L) = C_{n^* l^*}^2 f_{lm} I_p \sqrt{\frac{6}{\pi}}\left( \frac{2E_0}{E_L} \right)^{2n^*-|m|-3/2}
\times \exp{\left( - \frac{2E_0}{3 E_L}  \right)},
\end{equation}

We explicitly emphasize the use of the full $W_{PPT}$ cycle averaged ionization rate Eq. \ref{PPT_ionization_correct}, omitting the ADK approximation, for single electron ionization in this work.

For argon gas with an orbital angular momentum quantum number ($l=1$), we calculated the ionization rate by averaging over all possible values of its magnetic quantum number ($m$), representing the projection of the angular momentum along the direction of the applied field.  This methodology follows the approach outlined by Larochelle et al. in their investigation of the Coulomb effect in multiphoton ionization of rare-gas atoms \cite{Larochelle1_Coulomb_ionization_J_Phys_B_1998} and is consistent with the averaging procedure detailed in equation (61) of  PPT model \cite{Perelomov_PPT_JETP1966}.

\begin{equation}
\label{average_ionization_rate}
W_{PPT}(E_L,\omega_L) = \frac{1}{2l+1}\sum_{m=-l}^{l}W_{mPPT}(E_L,\omega_L)
\end{equation}


\subsection{Barrier-suppression ionization correction}

 The BSI correction to the ionization rate, described in  the Eq. \ref{BSI_ionization_rate} \cite{Zhang_Empirical_formula_BSI_PRA_2014}, has been implemented. The relevant parameter values $a_1, a_2$, and $a_3$ are provided in Table \ref{BSI_table}. It has been demonstrated that this modified ionization rate formula aligns well with $E_L \leq 4.5E_{BSI}$, falling within the range the peak laser field used in this work.

\begin{equation}
\label{BSI_ionization_rate}
W(E_L,\omega_L) = W_{PPT}(E_L,\omega_L)\exp\left(a_1\frac{E_L^2}{E_{BSI}^2}+a_2\frac{E_L}{E_{BSI}}+a_3 \right).
\end{equation}
(Note, that we corrected a typo in the the original paper \cite{Zhang_Empirical_formula_BSI_PRA_2014} , where a minus sign appears in the Eq. (8).)

\begin{table}[h]
\caption{Relevant values of parameters for BSI correction for our experiment \cite{Zhang_Empirical_formula_BSI_PRA_2014}.}
\begin{center}
\begin{tabular}{| c | c c | c c | c c |}
\hline
Parameter & & He & & Ar & & $\mathrm{Ar^{+1}}$ \\
\hline
$a_1$ & &  0.13550  & &  0.16178  & &  0.30441  \\
\hline
$a_2$ & & -0.86210  & &  -1.50441  & &  -2.70461  \\
\hline
$a_3$ & &  0.021562  & &  0.32127  & &  1.40821  \\
\hline
\end{tabular}
\label{BSI_table}
\end{center}
\end{table}

The BSI formula we use is a further development of the correction proposed by Tong et al. \cite{Tong_formula_BSI_JPhysB_2005}. It has been noted that their original correction for the tunneling ionization fits well for $E_L < 2.5E_{BSI}$. However, for larger values of $E_L$, the ionization rate decreases for argon, which is inconsistent with the expected physical behavior.

\subsection{Non-sequential double ionization}

Non-sequential Double Ionization (NSDI) \cite{Corkum_3step_PRL1993, Walker_double_ion_PRL_1994} is a process in which a singly charged ion undergoes secondary ionization due to recollision with the electron released earlier in the same laser cycle. 

The initial step of non-sequential double ionization (NSDI) involves the ionization of an atom within the laser cycle. Consequently, the instantaneous ionization rate \cite{Perelomov_PPT_JETP1966} is used:

\begin{equation}
\label{inst_ionization_rate}
W(E(t)) = I_p C_{n^*l^*}^2f_{lm}\left(\frac{2E_0}{E(t)} \right)^{2n^*-|m|-1}\exp\left\lbrace -\frac{2E_0}{3E(t)} \right\rbrace,
\end{equation}
where time dependency comes from the field, $E(t) = A(t)\cos(\omega t)$, where  $A(t)$ is the amplitude of the specific cycle under the slowly varying envelope approximation.


As discussed in the Main section of the manuscript, we calculate the probability of ionization, $P(t_i)$, in the time window $[t_i \; t_i+dt]$ within an optical cycle of period T. 

Electrons which return to their parent ion possess a probability of either ionization or excitation, determined by the electron's kinetic energy and the respective energy dependent cross-section for ionization/excitation by collision ($\sigma_{NSDI/NSE}$) \cite{Tawara_X_section_1987, Jesus_Atomic_JOSA_B_2004}. This probability is computed as an overlap between the electron's wavepacket spread in the laser field and the corresponding cross-section. The calculations are outlined as follows:

\begin{align}
P_c^{NSDI/NSE}(t_i) &= 2\pi \int_0^{R_c^{NSDI/NSE}(E_k(t_i))} |\Psi(r_\perp)|^2 r_\perp dr_\perp \nonumber \\ 
&= 2\pi N \int_0^{R_c^{NSDI/NSE}(E_k(t_i))} \exp\left(-\frac{r_\perp^2}{2\Delta r_\perp^2(t_i)} \right) r_\perp dr_\perp \nonumber \\
&=1-\exp\left( - \frac{\left(R_c^{NSDI/NSE}(E_k(t_i))\right)^2}{2\Delta r_\perp^2 (t_i)}\right)
\end{align}
Where, $N$ is the normalizing factor, 
\begin{equation}
N = \frac{1}{2\pi \Delta r_\perp^2}.
\end{equation}

$\Delta r_\perp$ is the electron wavepacket spread,
\begin{equation}
\Delta r_\perp(t_i) = \Delta p_\perp (t_r-t_i) = \frac{1}{\sqrt{2T_{tunnel}}} (t_r-t_i) = \frac{\sqrt{E(t_i)}}{(8I_p)^{1/4}}(t_r - t_i).
\end{equation}

$R_c$ is the collision radius,
\begin{equation}
R_c^{NSDI/NSE} = \sqrt{\frac{\sigma_{NSDI/NSE}(E_k(t_i))}{\pi}}.
\end{equation}

Finally, the NSDI rate and NSE rate are,

\begin{equation}
W_{NSDI/NSE} = \frac{\sum_{t_i}P(t_i)P_c^{NSDI/NSE}(t_i)}{\frac{T}{2}}.
\end{equation}

If the electron, following its collision with the parent ion, excites rather than ionizes the ion, the ion itself may undergo ionization from the laser field while in its excited state. This phenomenon is termed recollision-induced excitation with subsequent ionization (RESI) \cite{Kopold_RESI_PRL_2000, Feuerstein_RESI_PRL_2001, Bergues_NSDI_NatComm_2012}. From the ion's excited state, its ionization rate is calculated by Eq. \ref{BSI_ionization_rate}, with the proper values for the ion's $I_p$ and $Z=2$.

\subsection{Procedure for pulse duration extraction by fitting to the ionization curve}

 In our experiment, we measure the laser's pulse energy  ($\textrm{PE}$), the normalized spatial intensity distribution in the interaction region ($I_N(\vec{r}_\perp)$) as illustrated in Fig. \ref{Fig_BeamFocusProfile}, and the ion signal as a function of $\textrm{PE}$ (ionization curve). 

The $\textrm{PE}$ is related to the laser field amplitude $E(\vec{r}_\bot,t)$ by:

\begin{equation}
\label{E_field}
E(\vec{r}_\bot,t)=E_{peak} \sqrt{I_N(\vec{r}_\bot)} f(t),
\end{equation}
where $E_{peak}$ denotes the laser's pulse peak field. The pulse peak intensity ($I_{peak}$) can be expressed as:

\begin{equation}
\label{E_L2I_peak}
I_{peak} = c \varepsilon_0 E^2_{peak}/2.
\end{equation}

We then relate the peak intensity ($I_{peak}$) to the pulse energy ($\textrm{PE}$) using the equation:

\begin{equation}
\label{I_peak}
I_{peak} = \frac{\mathrm{PE}}{\iint_0^\infty I_N(\vec{r}_\perp)d^2\vec{r}_\perp \int_{-\infty}^\infty f^2(t)dt}.
\end{equation}
Here, $c$ is the speed of light in vacuum, $\varepsilon_0$ is the vacuum permittivity. Note that  $\int_{-\infty}^\infty f^2(t)dt = C_S \times \tau_{FWHM}$, where $C_S$ is a constant depending on the normalized temporal pulse envelope $f(t)$ (e.g., for $f(t)=\text{sech}\left(2\textrm{arcsech}(\sqrt{0.5})t/\tau_{FWHM}\right)$, $C_S=1/{\textrm{arcosh}\left(\sqrt{2}\right)}$).
Therefore, the pulse duration $\tau_{FWHM}$ is the only free fitting parameter. 

For each $\tau_{FWHM}$, we solve the set of ionization dynamics rate equations and calculate the total ion signal $Y_{ions}$ (see details in the main paper). We use the Matlab Curve Fitting Toolbox to fit the ionization curve to the calculated $Y_{ions}$ and determine the optimal $\tau_{FWHM}$, the fitting error, and the goodness of fit $R^2$.  Subsequently, we use Eq. \ref{I_peak} to extract the $I_{peak}$.

\clearpage 
\bibliography{G:/BGU_since_2013/papers/Atto_references_since_2024_10_11}

\section{Acknowledgements}
Acknowledgements: E.F. acknowledges the support of Israel Science Foundation (ISF) grant 2855/21. N.S. acknowledges financial support from the Kreitman Hi-tech student fellowship.

\section{Author Contributions}
E.F. conceived and supervised the project. N.S. carried out the measurements. All authors designed and built the experimental setup, performed the theoretical calculations and analysis,  interpreted and discussed the experimental and theoretical results, and contributed to the final manuscript.

\section{Conflict of Interest statement}

The authors declare no competing interests.

\end{document}